\documentclass[useAMS,usenatbib]{mn2e}

\usepackage{epsfig}
\usepackage{amsmath}
\usepackage{amssymb}
\usepackage{graphicx}
\usepackage{multirow}
\usepackage{relsize}
\usepackage{color}

\newcommand{\apj}{ApJ} 
\newcommand{\aj}{AJ} 
\newcommand{\apjl}{ApJ} 
\newcommand{\apjs}{ApJ} 
\newcommand{\aap}{A\&A} 
\newcommand{\araa}{ARAA} 
\newcommand{\mnras}{MNRAS} 
\newcommand{\nat}{Nature} 

\newcommand{\sigsfr}{\Sigma_\mathrm{SFR}}
\newcommand{\sigsf}{\Sigma_\mathrm{SF}}
\newcommand{\siggas}{\Sigma_\mathrm{gas}}
\newcommand{\ak}{A_\mathrm{K}}
\newcommand{\epssf}{\eps_\mathrm{SF}}
\newcommand{\epssfzero}{\eps_\mathrm{SF,0}}
\newcommand{\eps}{\epsilon}
\newcommand{\alphavir}{\alpha_\mathrm{vir}}
\newcommand{\sfe}{\mathrm{SFE}}
\newcommand{\sigs}{\sigma_s}
\newcommand{\phit}{\phi_t}
\newcommand{\tff}{t_\mathrm{ff}}
\newcommand{\mach}{\mathcal{M}}

\newcommand{\cs}{c_\mathrm{s}}

\newcommand{\cm}{\mathrm{cm}}
\newcommand{\km}{\mathrm{km}}
\newcommand{\pc}{\mathrm{pc}}
\newcommand{\s}{\mathrm{s}}
\newcommand{\yr}{\mathrm{yr}}
\newcommand{\Gauss}{\mathrm{G}}

\title[Variations in the star formation law]{The origin of physical variations in the star formation law} 

\author[Federrath]{Christoph~Federrath$^{1}$
\thanks{E-mail: christoph.federrath@monash.edu}\\
$^{1}$Monash Centre for Astrophysics, School of Mathematical Sciences, Monash University, VIC 3800, Australia}

\begin{document}



\maketitle

\begin{abstract}
Observations of external galaxies and of local star-forming clouds in the Milky Way have suggested a variety of star formation laws, i.e.~simple direct relations between the column density of star formation ($\sigsfr$: the amount of gas forming stars per unit area and time) and the column density of available gas ($\siggas$). Extending previous studies, we show that these different, sometimes contradictory relations for Milky Way clouds, nearby galaxies, and high-redshift discs and starbursts can be combined in one universal star formation law in which $\sigsfr$ is about 1\% of the local gas collapse rate, $\siggas/\tff$, but a significant scatter remains in this relation. Using computer simulations and theoretical models, we find that the observed scatter may be primarily controlled by physical variations in the Mach number of the turbulence and by differences in the star formation efficiency. Secondary variations can be induced by changes in the virial parameter, turbulent driving and magnetic field. The predictions of our models are testable with observations that constrain both the Mach number and the star formation efficiency in Milky Way clouds, external disc and starburst galaxies at low and high redshift. We also find that reduced telescope resolution does not strongly affect such measurements when $\sigsfr$ is plotted against $\siggas/\tff$.
\end{abstract}


\section{Introduction}

Stars form in dense molecular cores inside giant molecular clouds in the interstellar medium \citep{Ferriere2001}. These clouds are highly turbulent and magnetized, and are in approximate virial equilibrium with comparable values of the gravitational, kinetic and magnetic energy \citep{StahlerPalla2004}. Despite continuous efforts over the last decades, we still do not know which physical processes determine the star formation rate (SFR) in our Galaxy and in extragalactic systems, such as disc and starburst galaxies. We do know, however, that turbulence plays a key role in controlling the star formation process \citep{MacLowKlessen2004,ElmegreenScalo2004,McKeeOstriker2007}.
Almost all of our current knowledge about star formation comes from sub-millimetre observations. These observations provide us with maps of gas or dust column density ($\siggas$), which can be combined with young stellar object (YSO) counts, infrared, or ultraviolet luminosities, to yield the column density of star formation ($\sigsfr$). Such data have been collected for nearby and distant galaxies, and for clouds in the Milky Way (MW).

\begin{figure}
\centerline{\includegraphics[width=1.0\linewidth]{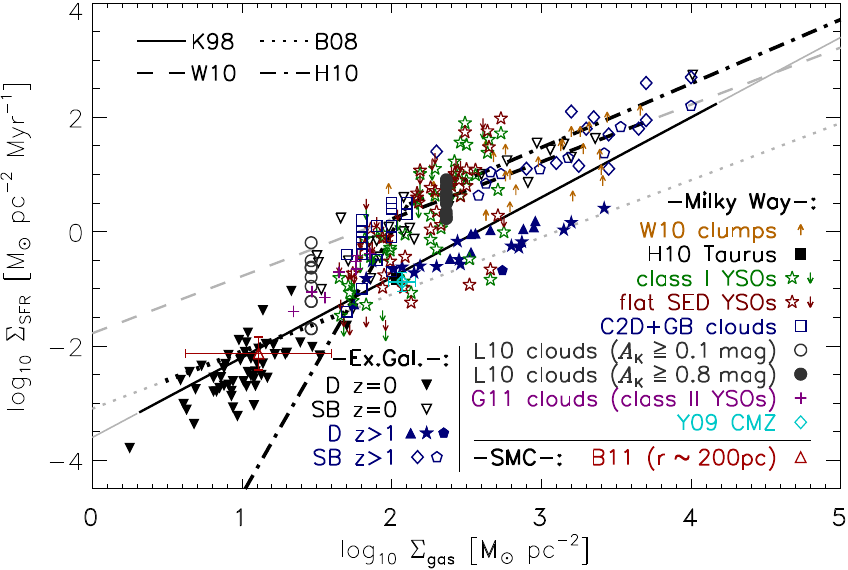}}
\caption{
Star formation rate column density ($\sigsfr$) versus gas column density ($\siggas$), measured in MW clouds, as well as in nearby and high-redshift disc and starburst galaxies. The data shown are from W10 \citep{WuEtAl2010}: HCN(1--0) clumps (uparrows); from H10 \citep{HeidermanEtAl2010}: Taurus (filled square), class I YSOs and flat SED YSOs (green and red stars with upper limits shown as downarrows), and C2D+GB clouds (open squares); from L10 \citep{LadaLombardiAlves2010}: molecular clouds observed at two different extinction thresholds ($\ak\ge0.1\,\mathrm{mag}$: open circles, and $\ak\ge0.8\,\mathrm{mag}$: filled circles); from G11 \citep{GutermuthEtAl2011}: class II YSO counts in eight molecular clouds (crosses); from Y09 \citep{YusefZadehEtAl2009}: the Central Molecular Zone (CMZ, turquoise diamond with error bars); and from B11 \citep{BolattoEtAl2011}: the Small Magellanic Cloud (SMC, red triangle with error bars).
Extragalactic data \citep{Kennicutt1998,BoucheEtAl2007,DaddiEtAl2010b,GenzelEtAl2010,TacconiEtAl2010} of disc (D) and starburst (SB) galaxies at low redshift ($z=0$) and high redshift (\mbox{$z\sim1$--$3$}) are reproduced from the tabulated compilation in KDM12 \citep{KrumholzDekelMcKee2012}.
(Table~3 in KDM12 for Ds and SBs contains naming errors and SB galaxy VII Zw 31, erroneously called `NGC 7552', has wrong $\siggas$ and $\sigsfr$. An erratum is in preparation [M.~Krumholz, private communication] and those errors have been corrected here.)
Typical uncertainties for the Ds and SBs are of the order of $0.5\,\mathrm{dex}$ (factor of 3) in both $\siggas$ and $\sigsfr$ \citep{Kennicutt1998}, but there may be additional uncertainties due to calibration errors caused by different forms of the initial mass function and different $\mathrm{CO/H}_2$ conversion factors \citep{DaddiEtAl2010b}.
Previously suggested star formation laws from extragalactic observations by K98 \citep{Kennicutt1998} and B08 \citep{BigielEtAl2008}, as well as from MW observations by W10 and H10 are shown as lines for comparison.
}
\label{fig:sfrcoldens}
\end{figure}

Figure~\ref{fig:sfrcoldens} shows a plot of $\sigsfr$ versus $\siggas$, combining the most recent measurements in MW clouds, as well as nearby and high-redshift disc and starburst galaxies. For comparison, four previously suggested star formation laws are shown with different line styles. First of all, we see that most of the MW data lie systematically above the extragalactic relations \citep{Kennicutt1998,BigielEtAl2008} by about an order of magnitude in $\sigsfr$. Secondly, for any given $\siggas$, we see a large range of $\sigsfr$, spanning about two orders of magnitude or more. Thirdly, the \citet[][hereafter L10]{LadaLombardiAlves2010} clouds, measured at an extinction threshold of $\ak\ge0.8\,\mathrm{mag}$ (filled circles) are systematically higher in both $\siggas$ and $\sigsfr$ than \emph{the same clouds} evaluated for $\ak\ge0.1\,\mathrm{mag}$ (open circles).
Given the broad distribution of observational data in Figure~\ref{fig:sfrcoldens}, a universal star formation law seems quite elusive. Although the overall correlation between $\sigsfr$ and $\siggas$ suggests that denser gas forms stars at a higher rate, the scatter is significant and there appears to be a bimodal distribution between disc and starburst galaxies.

Recently, \citet[][hereafter H10]{HeidermanEtAl2010} explained the systematic elevation of MW clouds over extragalactic systems by the fact that observations of star formation in MW clouds resolve individual sites of star formation, while observations of distant galaxies inevitably average over large areas, because of the limited telescope resolution. This alone, however, does not explain the bimodal distribution between disc and starburst galaxies seen in Figure~\ref{fig:sfrcoldens}.

\section{A more universal star formation law}

\begin{figure*}
\centerline{\includegraphics[width=1.0\linewidth]{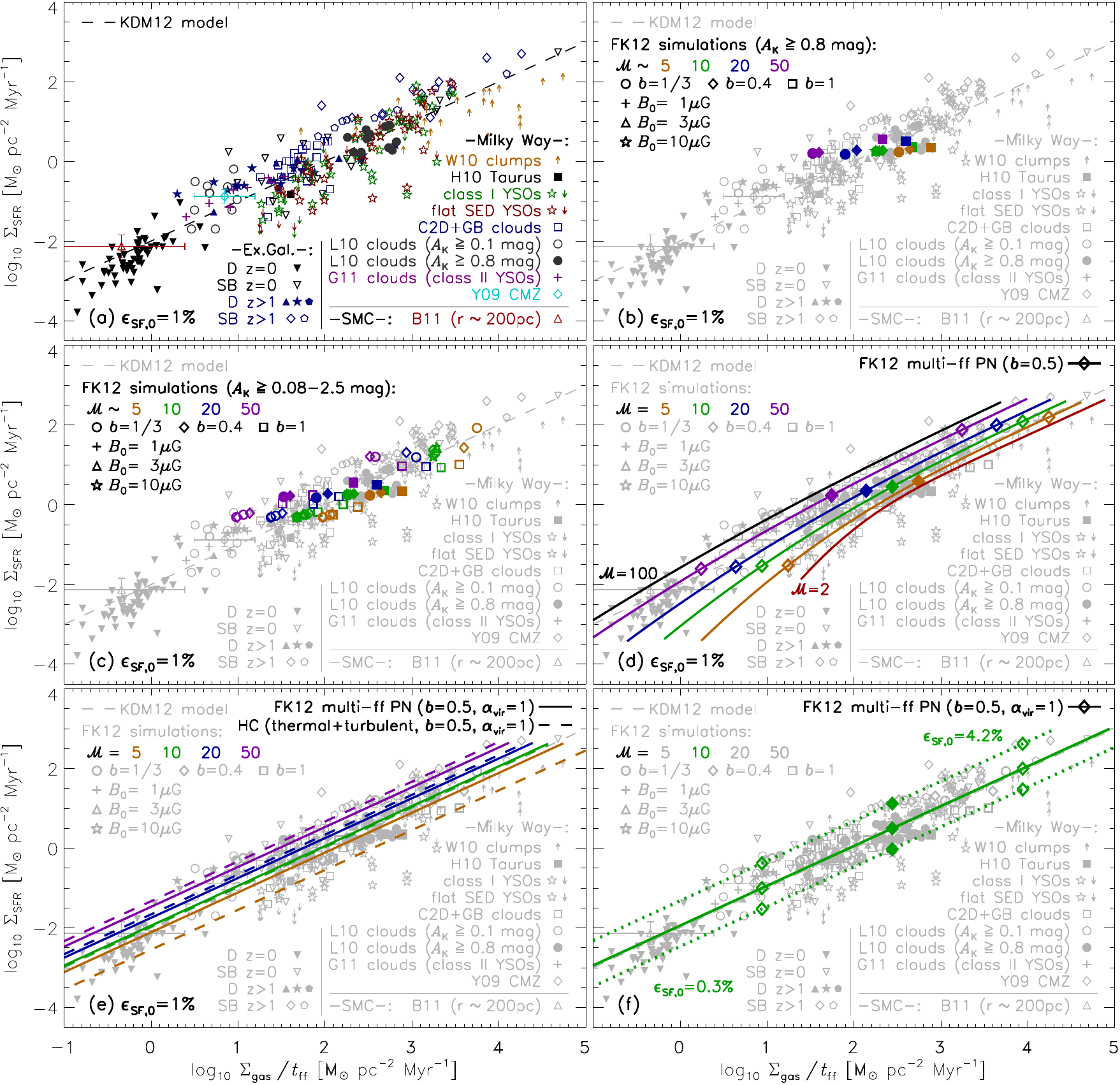}}
\caption{
\textbf{(a)}: Same as Figure~\ref{fig:sfrcoldens}, but showing $\sigsfr$ as a function of $\siggas/\tff$. Equation~(\ref{eq:kdm}) with $\epssfzero=1\%$ suggested by KDM12 is shown as the dashed line. Observational uncertainties (see Figure~\ref{fig:sfrcoldens} caption) are higher in this representation, because of the additional uncertainties in $\tff$ (see Appendix~\ref{app:caveats}).
\textbf{(b)}: Same as (a), but with simulations from FK12 \citep{FederrathKlessen2012} evaluated for $\epssfzero=1\%$ and $\ak\ge0.8\,\mathrm{mag}$ superposed. These are hydrodynamic computer models with turbulent Mach number $\mach\sim5$ (orange), $10$ (green), $20$ (blue), and $50$ (purple), respectively, for solenoidal ($b=1/3$; circles), mixed ($b=0.4$; diamonds), and compressive driving ($b=1$; squares) of the turbulence. Magnetohydrodynamic simulations with $\mach\sim10$, $b=0.4$, and typical magnetic fields, $B_0=1$, $3$, and $10\,\mu\Gauss$ (cross, triangle, and star) are also shown, but these magnetized models are almost indistinguishable from the $B=0$ models.
\textbf{(c)}: Same as (b), but the simulations were not only evaluated at $\ak\ge0.8\,\mathrm{mag}$ (filled symbols), but at a range of extinction thresholds, \mbox{$\ak\ge0.08$--$2.5\,\mathrm{mag}$} (open symbols).
\textbf{(d)}: Same as (c), but with theoretical model curves
given by Equations~(\ref{eq:sigsfr}) and~(\ref{eq:epssf}) superposed for a typical turbulent driving parameter $b=0.5$ and $\mach=2$, $5$, $10$, $20$, $50$, and $100$, each evaluated at different densities (equivalent to a range of extinction thresholds, to cover the range of observational $\siggas/\tff$).
\textbf{(e)}: Same as (d), but enforcing a virial parameter $\alphavir=1$ in Equation~(\ref{eq:epssf}) and additionally showing the \citet{HennebelleChabrier2011} (HC) version of our model for $\epssf$ as the dashed lines.
\textbf{(f)}: Same as (e), but showing the effect of varying the efficiency \mbox{$\epssfzero=0.3\%$--$4.2\%$} in Equation~(\ref{eq:epssf}) for fixed $\mach=10$.
}
\label{fig:sfrcoldens_kdm}
\end{figure*}

More recently, \citet[][hereafter KDM12]{KrumholzDekelMcKee2012} thus argued that the standard star formation relation shown in Figure~\ref{fig:sfrcoldens} may not provide the best physical representation. Based on the assumption that the SFR is inversely proportional to the dynamical time of the gas \citep{Schmidt1959,Elmegreen2002}, KDM12 suggest that a better fit is obtained, if $\sigsfr$ is plotted against $\siggas/\tff$, i.e.~$\sigsfr$ as a function of $\siggas$ divided by the local gas collapse time,
\begin{equation} \label{eq:tff}
\tff(\rho)=\left(\frac{3\pi}{32\,G\rho}\right)^{1/2},
\end{equation}

evaluated for each cloud or galactic system individually. Although not directly observable, the gas density $\rho=(3\sqrt{\pi}/4)M/A^{3/2}$ with the cloud mass $M=\siggas A$ and the observed area $A$ can be estimated by assuming that the clouds are approximately spherical objects (KDM12), introducing additional uncertainties (Appendix~\ref{app:caveats}). For extragalactic systems, the gas collapse time is taken to be the minimum of the Toomre time for stability of the disc or starburst and the local cloud freefall time (see KDM12 for details\footnote{The high-z disc and starburst galaxy dataset in table~4 of KDM12 contains errors related to the computation of the Toomre time [M.~Krumholz, private communication]. Figure~\ref{fig:sfrcoldens_kdm} here shows the corrected data.}). In this way, the MW clouds and the extragalactic data seem to exhibit a much tighter correlation, which is shown in Figure~\ref{fig:sfrcoldens_kdm}(a). KDM12 only included the C2D+GB clouds from H10 and the L10 clouds at the two different extinction thresholds, while here we add all data from H10, \citet[][hereafter W10]{WuEtAl2010}, and the clouds observed in \citet[][hereafter G11]{GutermuthEtAl2011}. We also include an average of the $200\,\pc$ resolution data ($A=4.5\times10^4\,\pc^2$; A.~Bolatto, private communication) of the Small Magellanic Cloud (SMC) \citep{BolattoEtAl2011}.
One might question whether mixing resolved measurements of MW clouds and galactic discs with unresolved discs and starbursts (KDM12) in a single plot produces a physically meaningful comparison, because of extinction and telescope resolution issues \citep[e.g.][]{CalzettiLiuKoda2012,ShettyKellyBigiel2013}. Encouragingly, however, we find in tests with synthetic observations at different extinction thresholds and telescope resolutions varying by a factor of 32 that measurements presented in the form of Figure~\ref{fig:sfrcoldens_kdm} vary by less than a factor of two for fixed physical conditions (Appendix~\ref{app:telescope}).

The dashed line in Figure~\ref{fig:sfrcoldens_kdm}(a) shows the empirical relation by KDM12,
\begin{equation} \label{eq:kdm}
\sigsfr=\epssfzero\times\siggas/\tff\,,
\end{equation}
with a constant proportionality factor, $\epssfzero=1\%$, which we define here as the total star formation efficiency, $\epssfzero\equiv\eps\times\sfe$. In this expression for $\epssfzero$, the \emph{local} core-to-star efficiency, \mbox{$\eps=0.3$--$0.7$}, is the fraction of infalling gas that is accreted by the star \citep{MatznerMcKee2000}, i.e.~about half. The other half is expelled by jets, winds and outflows. The \emph{global} (cloud-scale) efficiency, \mbox{$\sfe=1\%$--$6\%$}, is the typical fraction of gas forming stars in a whole molecular cloud \citep{EvansEtAl2009,LadaLombardiAlves2010,FederrathKlessen2013}. This yields a combined, total star formation efficiency, \mbox{$\epssfzero\sim0.3\%$--$4.2\%$}. Here we adopt an intermediate value, $\epssfzero=1\%$, as favoured in observations and analytic models \citep{KrumholzTan2007,RenaudEtAl2012}; however, we also study the influence of varying $\epssfzero$ below.
The observational data in Figure~\ref{fig:sfrcoldens_kdm}(a) indeed exhibit a better correlation than in Figure~\ref{fig:sfrcoldens}, yet the scatter is still significant and remained largely unexplained in KDM12.
What is the origin of this persistent scatter?

To advance on this issue, we compare the observations with computer simulations from \citet[][hereafter FK12]{FederrathKlessen2012}, covering a substantial range of observed physical cloud parameters with Mach numbers $\mach=\sigma_v/\cs=5$--$50$, different driving of the turbulence parametrized by $b=1/3$ for solenoidal (divergence-free), $b=0.4$ for mixed and $b=1$ for compressive (rotation-free) driving, as well as a few different magnetic field strengths with $B=1$, $3$ and $10\,\mu\Gauss$ for $\mach=10$ cases.
In Figure~\ref{fig:sfrcoldens_kdm}(b), we superpose these computer simulations, measured at a fixed extinction threshold, $\ak\ge0.8\,\mathrm{mag}$ and for $\epssfzero=1\%$. To do this, we first produce column density projections along each coordinate axis of the three-dimensional simulations, when $\sfe=2\%$ of the total cloud mass has been accreted by sink particles \citep{FederrathBanerjeeClarkKlessen2010}. Multiplying this by the core-to-star efficiency $\eps=0.5$ yields the target $\epssfzero=1\%$ as for the KMD12 model. We then measure $\siggas$ in structures above a given $\ak$ threshold in each projection and determine the amount of gas that formed sink particles, $\sigsf$, in the corresponding $\ak$ contour\footnote{We do not distinguish connected from disconnected structures. We simply take a column density threshold and sum up all the simulation pixels that are above a given $\ak$ threshold.}. We take the total mass in gas and the total mass in sink particles above a given $\ak$ threshold and divide both by the cloud area that is above that extinction threshold to measure $\siggas$ and $\sigsf$, respectively. Finally, we compute $\sigsfr=\sigsf/(2\,\mathrm{M}\yr)$ for class II YSOs, routinely applied by observers \citep{EvansEtAl2009,HeidermanEtAl2010,LadaLombardiAlves2010}, such that our procedure to place the simulation data in Figure~\ref{fig:sfrcoldens_kdm}(b) matches the observational method as closely as possible. Note that this procedure does not necessarily reflect the true rate of star formation in the simulations (studied in detail in FK12), but places the simulation data as they would be placed if processed by an observer, who does not have any information about the time evolution.

Figure~\ref{fig:sfrcoldens_kdm}(b) shows that the simulations are consistent with the observations and roughly agree with the L10 clouds measured at the same extinction threshold, $\ak\ge0.8\,\mathrm{mag}$. Comparing the simulations with one another, we arrive at three conclusions. First, for a fixed Mach number, the simulations with compressive driving exhibit higher $\siggas/\tff$ and $\sigsfr$ than the respective simulations with mixed and solenoidal driving. Secondly, $\siggas/\tff$ decreases with increasing $\mach$, while $\sigsfr$ stays almost constant. Thirdly, magnetic fields reduce $\siggas/\tff$, but only very marginally.

Evaluating the same simulations as in Figure~\ref{fig:sfrcoldens_kdm}(b) not only at $\ak\ge0.8\,\mathrm{mag}$, but at a range of extinction thresholds, \mbox{$\ak\ge0.08$--$2.5\,\mathrm{mag}$}, we obtain the distribution of simulation data shown in Figure~\ref{fig:sfrcoldens_kdm}(c). We find that the roughly linear proportionality between $\sigsfr$ and $\siggas/\tff$ is primarily driven by changes in the extinction value defining the clouds. This was already seen when we compared the L10 clouds at $\ak\ge0.1\,\mathrm{mag}$ and $\ak\ge0.8\,\mathrm{mag}$ in panel (a).
The only difference is that the simulation data do not have the dynamic range (because of limited numerical resolution) to reach down to the very low extinction values in the L10 clouds.

Figure~\ref{fig:sfrcoldens_kdm}(c) confirms the effect of increasing sonic Mach number seen in panel (b), i.e.~clouds with higher $\mach$ shift to lower $\siggas/\tff$. The reason for this is that $\siggas$ is almost fixed for a given $\ak$ threshold, but $1/\tff\propto\rho^{1/2}\propto \ell^{-1/2}\propto \mach^{-1}$ varies with cloud size $\ell$ and Mach number (which is why clouds with Mach 50 have about $10\times$ lower $\siggas/\tff$ than Mach 5 clouds in Figure~\ref{fig:sfrcoldens_kdm}b) as the simulations roughly follow the \citet{Larson1981} relations for the velocity dispersion and density as a function of cloud size (see FK12, table~2). A substantial fraction of the observed scatter in $\siggas/\tff$ may thus be explained by variations in the turbulent Mach number. If the clouds that form stars actually follow the Larson relations in the same way as the simulations do here, then we may relate the observational data directly to the simulation data. However, some regions do not follow the standard Larson relations \citep[e.g.~the Central Molecular Zone (CMZ); see][and potentially also extragalactic systems]{ShettyEtAl2012}, such that those regions will probably not be consistent with the simulations. However, the scatter seen in the observations may still be attributable to variations in the Larson relations.

\section{A theoretical model for $\sigsfr$}

To substantiate this finding, we add theoretical model curves in Figure~\ref{fig:sfrcoldens_kdm}(d). These models are based on the statistics of supersonic magnetohydrodynamic turbulence in self-gravitating systems \citep{KrumholzMcKee2005,PadoanNordlund2011,HennebelleChabrier2011}. Here we focus on the best-fitting multi-freefall PN model in FK12 and compute
\begin{equation} \label{eq:sigsfr}
\sigsfr=\epssf\times\siggas/\tff,
\end{equation}
where $\siggas=\rho\ell$, i.e.~the product of gas density $\rho$ and size $\ell$ of the cloud structure. Equation~(\ref{eq:sigsfr}) is the same as Equation~(\ref{eq:kdm}), but instead of a constant proportionality factor $\epssfzero$, we evaluate the dimensionless function
\begin{equation} \label{eq:epssf}
\epssf=\frac{\epssfzero}{2\phit}\exp\left(\frac{3}{8}\sigs^2\right)\left[1+\mathrm{erf}\left(\frac{\sigs^2-s_\mathrm{crit}}{\sqrt{2\sigs^2}}\right)\right].
\end{equation}
Equation~(\ref{eq:epssf}) is derived from an integral over the high-density tail of the log-normal probability distribution function (PDF) of the turbulent gas density \citep{Vazquez1994,FederrathKlessenSchmidt2008}\footnote{Although the PDF can develop a power-law tail when gas starts to collapse \citep{Klessen2000,CollinsEtAl2012}, a strong tail only occurs once $\sfe\approx5\%$ \citep{FederrathKlessen2013}, at which point star formation typically shuts off due to feedback processes. We thus conclude that a log-normal PDF is a reasonably good approximation for a simple theoretical model of the SFR, even when the density structure comes from a mixture of turbulence, gravitational instabilities, feedback, or cooling and heating processes \citep{WadaNorman2001,BournaudEtAl2010,GloverFederrathMacLowKlessen2010}.},
\begin{equation}
p(s)=\frac{1}{\sqrt{2\pi\sigs^2}}\exp\left(-\frac{(s-s_0)^2}{2\sigs^2}\right)\,,
\label{eq:pdf}
\end{equation}
expressed in terms of the logarithmic density, \mbox{$s\equiv\ln{(\rho/\rho_0)}$}, where $\rho_0$ is the mean density and $s_0=-0.5\,\sigs^2$ is the logarithmic mean density. This integral is weighted by $\rho/\rho_0$ to estimate the mass fraction of gas above a critical density $s_\mathrm{crit}$ and weighted by a freefall-time factor to construct a dimensionless SFR:
\begin{equation}
\epssf=\frac{\epssfzero}{\phit} \int_{s_\mathrm{crit}}^{\infty}{\frac{\tff(\rho_0)}{\tff(\rho)}\frac{\rho}{\rho_0}\,p(s)\mathrm{d}s}\,.
\label{eq:epssfbasic}
\end{equation}
Note that the factor $\tff(\rho_0)/\tff(\rho)$ is evaluated \emph{inside} the integral because gas with different densities has different freefall times \citep{HennebelleChabrier2011,HennebelleChabrier2013}.
The factor $\epssfzero$ is the same as in Equation~(\ref{eq:kdm}), and $1/\phit$ \citep[][]{KrumholzMcKee2005} accounts for the uncertainty in the timescale factor, which was measured to $1/\phit\approx0.5$ in FK12. 

The variables $\sigs$ and $s_\mathrm{crit}$ in Equation~(\ref{eq:epssf}) are the standard deviation of the density PDF \citep{MolinaEtAl2012},
$\sigs^2=\ln\left[1+b^2\mach^2\beta/(\beta+1)\right]$,
and the critical density \citep{PadoanNordlund2011},
$s_\mathrm{crit}=\ln{[0.067\theta^{-2}\alphavir\mach^2f(\beta)]}$
with $f(\beta)=(1+\beta^{-1})^{-2}(1+0.925\beta^{-3/2})^{2/3}$ and the virial parameter $\alphavir=5\sigma_v^2/(G\rho\ell^2)$ \citep{BertoldiMcKee1992}. The numerical factor $\theta\approx1$ was measured in FK12 and physically motivated in \citet{PadoanNordlund2011}. Combining all this yields $\epssf\equiv\epssf(\alphavir,\mach,b,\beta)$, i.e.~a dimensionless SFR as a function of four basic cloud parameters: $\alphavir$, $\mach$, the turbulent driving parameter $1/3\leq b\leq1$ \citep{FederrathKlessenSchmidt2008,FederrathDuvalKlessenSchmidtMacLow2010}, and the ratio of thermal to magnetic pressure, $\beta$. 

Since we concluded from the simulations above that magnetic fields only have a relatively weak effect \citep[with very strong magnetic fields, the SFR is reduced by a factor of 2--3, see][FK12]{PadoanNordlund2011,PadoanHaugboelleNordlund2012}, for simplicity we only consider theoretical cases without magnetic fields in the following ($\beta\to\infty$). Although there is no doubt that magnetic fields modify the picture, they are unlikely the primary controller of the order-of-magnitude variations that we see in the observations. For the same reason, we only consider a fixed, intermediate turbulent driving parameter $b=0.5$ \citep{Brunt2010,PriceFederrathBrunt2011,KainulainenTan2013,KainulainenFederrathHenning2013}. We further use the total efficiency $\epssfzero=1\%$ as before.

Figure~\ref{fig:sfrcoldens_kdm}(d) shows Equation~(\ref{eq:sigsfr}) evaluated for four cloud sizes $\ell=1$, $4$, $16$, and $100\,\pc$. These correspond to $\mach\sim5$, $10$, $20$, and $50$, according to the velocity dispersion--size relation  \citep{Larson1981,HeyerBrunt2004}, $\sigma_v=\mach\cs\approx1\,\km\,\s^{-1} (\ell/\pc)^{0.5}$ with $\cs\approx0.2\,\km\,\s^{-1}$, typical for molecular gas with temperatures of about \mbox{$10\,\mathrm{K}$} and standard solar composition \citep{OmukaiEtAl2005}. We also note in this context that \citet{Dib2011} and \citet{GloverClark2012} find that the SFR depends slightly on metallicity, introducing changes by a factor of 2--3, so the order-of-magnitude variations seen in observations cannot be explained by metallicity effects alone, but they may contribute.

In order to cover the range of $\ak$ and $\siggas/\tff$ in the observations, we vary the density along each theoretical model curve as a free parameter.
Using the density--size relation \citep{Larson1981,MacLowKlessen2004,McKeeOstriker2007} $\rho=\rho_0(\ell/\pc)^{-1}$ with a typical density scale $\rho_0=10^4\mu_\mathrm{H}\,\cm^{-3}$ (where $\mu_\mathrm{H}=1.67\times10^{-24}\,\mathrm{g}$ is the atomic mass of hydrogen), similar to the simulated clouds in Figure~\ref{fig:sfrcoldens_kdm}(b) and similar to the L10 clouds for $\ak\ge0.8\,\mathrm{mag}$, we obtain the filled diamonds in Figure~\ref{fig:sfrcoldens_kdm}(d), which agree well with the computer models for that extinction threshold. We also add the open diamonds, representing the same theoretical data, but for $10\times$ larger and smaller density scale $\rho_0$. For a given density scale, $\sigsfr$ is almost independent of $\mach$, only $\siggas/\tff\propto\mach^{-1}$ for constant $\siggas$ as we saw above for the simulation data. This implies that $\epssf$ in Equation~(\ref{eq:epssf}) increases with $\mach$ (because increasing $\mach$ leads to stronger gas compression and thus higher relative SFRs, see FK12), effectively compensating the decrease of $\siggas/\tff$ with $\mach$. Indeed, $\epssf\propto\mach^{3/4}$ for $\mach\gtrsim10$ and $\alphavir\sim1$, leading to a weak dependence of $\sigsfr\propto\mach^{-1/4}$ for fixed $\siggas$.

Changing the density scale $\rho_0$ in Figure~\ref{fig:sfrcoldens_kdm}(d) means that the virial parameter is about unity for the filled diamonds and about $0.1$ and $10$ for the open diamonds, respectively to the right and to the left of the filled diamonds ($\alphavir=1$). Such a systematic correlation of $\alphavir$ with $\sigsfr$ is rather unexpected, which is why we add another panel (e) where we keep $\alphavir=1$ in Equation~(\ref{eq:epssf}) along each model curve. Although variations in $\alphavir$ by at least two orders of magnitude are measured for MW clouds \citep{RomanDuvalEtAl2010,PadoanNordlund2011,KauffmannEtAl2013} and certainly contribute to the scatter, the overall Mach number dependence remains, even if we enforce $\alphavir=1$.

Figure~\ref{fig:sfrcoldens_kdm}(e) additionally shows the \citet{HennebelleChabrier2011,HennebelleChabrier2013} (HC) version of our model for $\epssf$ with otherwise identical parameters and $y_\mathrm{cut}=0.1$ (see FK12 for details of that model). The Mach number dependence is stronger, because the critical density in the HC model is $\rho_\mathrm{crit}\propto\mach^{-2}$ unlike in the PN model, where $\rho_\mathrm{crit}\propto\mach^2$ (see FK12, table~1). Both the multi-freefall PN and HC models support the basic idea that variations in the star formation relation may be caused by variations in the Mach number, but the details of that dependence are subject to significant uncertainties, introduced by the particular choice of model.

Finally, Figure~\ref{fig:sfrcoldens_kdm}(f) shows the effect of varying the efficiency \mbox{$\epssfzero=0.3\%$--$4.2\%$} in Equation~(\ref{eq:epssf}) for fixed $\mach=10$, which covers a substantial fraction of the observed variations in $\sigsfr$. Thus, for any point in the \mbox{$\sigsfr$--$\siggas/\tff$} relation, there is a degeneracy between the Mach number and the efficiency, which can only be broken by measuring both $\mach$ and $\epssfzero$ simultaneously.

\section{Discussion and Conclusion}

The main conclusion of this paper is that the observed scatter in the star formation law can be primarily explained by physical variations in the turbulent Mach number $\mach$ and the star formation efficiency $\epssfzero$. We find that the observed scatter is not random, but instead depends systematically on $\mach$ and $\epssfzero$. For a fixed extinction threshold or fixed $\siggas$, we find that $\siggas/\tff\propto\mach^{-1}$, if the standard Larson relations are in effect. Although some regions do not follow the standard Larson scalings (e.g.~the CMZ and possibly extragalactic regions), we still expect a variation of $\siggas/\tff$ also in such cases, albeit with a potentially different dependence. We further find that for fixed $\siggas$, the variations in $\sigsfr$ may be explained by variations in the star formation efficiency, systematically $\sigsfr\propto\epssfzero$ (see Figure~\ref{fig:sfrcoldens_kdm}f). The theoretical model, Equation~(\ref{eq:epssf}), also implies that some fraction of the scatter may be explained by variations in the virial parameter $\alphavir$, the turbulent driving parameter $b$, and the thermal-to-magnetic pressure ratio $\beta$.

We note that \citet{RenaudEtAl2012} have also recently developed an analytic model for $\sigsfr$ based on the log-normal density PDF and investigated the Mach number dependence of their model in the context of Kennicutt-Schmidt relations, such as plotted in Figure~\ref{fig:sfrcoldens}. They use Mach numbers in the range 1--20, gas scale heights of \mbox{$5$--$2000\,\pc$} and density thresholds of \mbox{$10$--$100\,\cm^{-3}$} to explain observations of MW clouds, discs and starbursts. The relatively low Mach numbers come about, because they chose to evaluate $\mach$ for temperatures of the warm interstellar medium ($T\approx10^{3-4}\,\mathrm{K}$). It is, however, the cold, molecular phase with $T\approx10^{1-2}\,\mathrm{K}$ \citep[where a log-normal PDF seems reasonable; see][]{GloverFederrathMacLowKlessen2010} in which stars form, so the relevant Mach numbers for the star-forming gas are about an order of magnitude higher than assumed in \citet{RenaudEtAl2012}, because $\mach\propto T^{-1/2}$.

Our simulations and theoretical models in Figure~\ref{fig:sfrcoldens_kdm} make direct predictions that can be tested with observations. If the Mach number and star formation efficiency were indeed the primary physical reasons for the variations in the star formation relation, then measuring $\mach$ and $\epssfzero$ in clouds and galaxies will eventually enable us to test these predictions. For example, the clouds and YSO data in the MW are in the expected range, \mbox{$\mach\sim2$--$20$} and \mbox{$\epssfzero\sim0.3\%$--$4.2\%$}, consistent with our theoretical models. The placement of the CMZ is also consistent with $\mach\sim50$, given the uncertainties in the data. However, measurements of $\mach$ and $\epssfzero$ in extragalactic systems are more difficult. For Arp~220, \mbox{$\mach\approx100$} with large uncertainties \citep{DownesSolomon1998}. Arp 220 (the rightmost downward pointing triangle in Figure~\ref{fig:sfrcoldens_kdm}(a)\footnote{The starburst galaxy Arp 220 is erroneously listed as `NGC 6946' in KDM12, because of a mismatch of the original K98 tables and the KDM12 table for disc and starburst galaxies. An erratum of KDM12 is in preparation (M.~Krumholz, private communication). A corrected table is available upon request.}) would be more consistent with $\mach\sim10$ for $\epssfzero=1\%$, which means that either our model is incorrect or $\epssfzero$ is relatively small for that galaxy, or the measurements of $\mach$, $\sigsfr$, $\siggas$ and $\tff$ are so uncertain for Arp 220 that it cannot be used to falsify the model, or the standard Larson relations do not apply for Arp 220, such that a direct matching of Mach numbers there and in our models (that assume standard Larson scaling) cannot be done with the present data. Finally, the SMC has velocity dispersions of $10$--$40\,\km\,\s^{-1}$ \citep{BekkiChiba2009}, which gives \mbox{$\mach=16$--$200$} for \mbox{$T=10$--$100\,\mathrm{K}$}, basically consistent with our theoretical model in Figure~\ref{fig:sfrcoldens_kdm}, but also with large uncertainties, so we need future observations that simultaneously constrain $\mach$ and $\epssfzero$.

\section*{Acknowledgements}
The author acknowledges comments by Frederic Bournaud, Neal Evans, Ralf Klessen, Mark Krumholz, Chris McKee, Daniel Price and Florent Renaud. We thank the anonymous referee for a critical and careful reading, which improved the paper significantly. This work was supported by ARC grant DP110102191. The simulations were run at LRZ (pr32lo) and JSC (hhd20).

\appendix

\section{Caveats and Limitations} \label{app:caveats}
Here we discuss caveats and limitations of the present study. First, unlike the classical Kennicutt-Schmidt relation, which only requires measurements of column-integrated quantities, $\sigsfr$ and $\siggas$, the KDM12 model requires an additional estimate of the volume density $\rho$ to compute the freefall time for the abscissa $\siggas/\tff$. The current estimate of $\rho$ by KDM12 assumes that the gas is homogeneously distributed along the line of sight (LOS). This is clearly an oversimplification, because the gas along the LOS has likely a range of densities and potentially contributions from different cloud components in the case of very long LOS. Eventually, a refined model would take the multi-freefall contributions of the PDF of gas densities along the LOS into account. Secondly, most of the MW cloud and YSO data use a fixed star formation time scale of $2\,\mathrm{M}\yr$ for the class II phase \citep{EvansEtAl2009,HeidermanEtAl2010,LadaLombardiAlves2010} to estimate $\sigsfr$. However, the exact value of $\sigsfr$ depends on the evolutionary phase and requires information about the time evolution of the cloud, which is not available from observations. Thus, estimates of $\sigsfr$ are highly uncertain and some spread of the data is likely caused by this effect \citep[for effects of different star formation timescales, see][]
{FederrathKlessen2012}.

\section{Effects of the telescope resolution} \label{app:telescope}

\begin{figure}
\centerline{\includegraphics[width=0.95\linewidth]{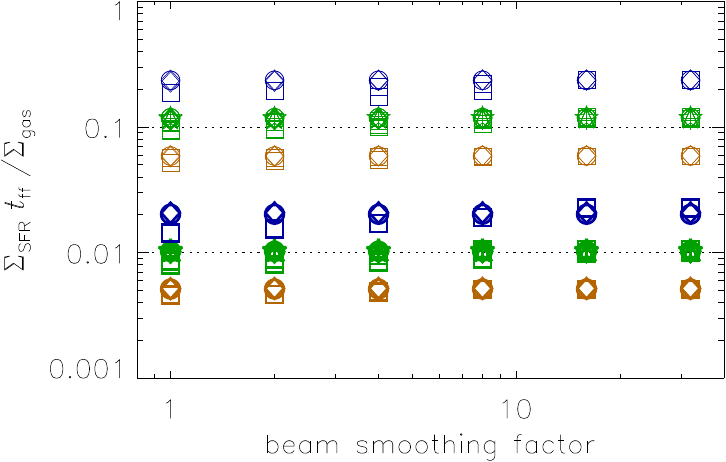}}
\caption{$\epssf=\sigsfr\tff/\siggas$ as a function of beam smoothing factor for $\epssfzero=1\%$ (lower set) and $\epssfzero=10\%$ (upper set). Varying the telescope resolution merely shifts the data along the star formation law, $\epssf=\mathrm{const}$ (shown as the dotted lines for $\epssfzero=1\%$ and $10\%$ for $\mach=10$ simulations; green symbols). Simulation symbols and colours are the same as in Figure~\ref{fig:sfrcoldens_kdm}(b)~and~(c).}
\label{fig:telescope}
\end{figure}

Figure~\ref{fig:telescope} shows the influence of the telescope resolution. We made synthetic observations of the simulations as in Figure~\ref{fig:sfrcoldens_kdm}(b), but with up to $32\times$ beam smoothing ($32\times$ reduced telescope resolution or observing the same cloud at a $32\times$ greater distance). Although $\sigsfr$ and $\siggas/\tff$ are both reduced by beam smoothing, they are reduced by roughly the same factor, such that $\epssf$ is almost independent of telescope resolution. This result is encouraging for observations, because it shows that $\epssf$ could be measured even with relatively low resolution.


\end{document}